# Transition from Integer Quantum Hall State to Insulator

X.C. Xie

*Department of Physics, Oklahoma State University, Stillwater, OK 74078.*

D.Z. Liu

*James Franck Institute, University of Chicago, Chicago, IL 60637*

(Submitted to Phys. Rev. B on November 9, 1995)

We study disorder induced transition between quantum Hall states and insulator state in a *lattice* model. We find the positions of the extended states do not change as disorder strength is varied. As a consequence, there are direct transitions from all Landau level quantum Hall states to insulator states, in contrast to the global phase diagram from early studies based on continuous models. We also provide the microscopic understanding of the transition in terms of the topological properties of the system.

It is well known that all electrons in a two-dimensional system are localized in the absence of magnetic field according to the scaling theory of localization [1]. When the two-dimensional electron system is subject to a strong perpendicular magnetic field, the energy spectrum becomes a series of impurity broadened Landau levels. The external magnetic field breaks the time-reversal symmetry and as a consequence, extended state appears in the center of each Landau band while states away from the Landau band centers are still localized. This gives rise to the integer quantum Hall effect [2]. Recently there have been much interest to understand the evolution of the extended states as the magnetic field goes to zero or equivalently as the disorder increases such that eventually all the extended states should disappear [3–7].

Earlier work by Laughlin [8] and Khmelnitskii [9] concluded that the extended state energy for the $n$th Landau level (LL) behaves in the following form:

$$E_n^c = (n+1/2)\hbar\omega_c \frac{1 + (\omega_c\tau)^2}{(\omega_c\tau)^2}, \qquad (1)$$

where $\omega_c$ is the cyclotron frequency and $\tau$ is impurity scattering time. According to this formula, $E_n^c$ deviates from the linear magnetic field dependence when $\omega_c\tau \sim 1$ and starts to float up as $B$ decreases or as disorder increases. Based on the floating up picture from continuous models by Laughlin and Khmelnitskii, Kivelson, Lee and Zhang [10] proposed the global phase diagram for transitions between the quantum Hall states (QHS) and insulator (or localized) states. There are two important consequences from the global phase diagram: (i) as disorder increases for a fixed field, no direct transition between higher LL ($n \neq 0$) QHS and insulator is allowed; (ii) as $B$ decreases for fixed disorder, reentry behavior should be observed in quantized Hall effect, *e.g.* $\nu = 1 \rightarrow \nu = 2 \rightarrow \nu = 1$.

In an earlier work [11] by us and Niu, we found that the floating-up picture is not valid in the *lattice* model. We concluded that: (1) the extended state energy $E_c$ for each Landau level is *always* linear in magnetic field; (2) for a given Landau level and disorder configuration there exists a critical magnetic field $B_c$ below which the extended state disappears; (3) the lower LLs are more robust to the metal-insulator transition with smaller $B_c$. We attributed the above results to strong LL coupling effect.

In this paper, we expand our study on disorder induced transition between quantum Hall states and insulator state in a lattice model. By carrying out finite-size scaling calculations of localization length, we demonstrate that the positions of the extended states do not change as disorder strength is varied. As a consequence, we show unambiguous evidence from our finite-size scaling analysis that there are direct transitions from all Landau level quantum Hall states to insulator states, in contrast to the global phase diagram [10]. We also provide the microscopic understanding of the transition in terms of the topological properties of the system.

In the following, we briefly outline our model [12,13] and technique. We model our two-dimensional system in a very long strip geometry with a finite width ($M$) square lattice with nearest neighbor hopping. Periodic boundary condition in the width direction is used to get rid of the edge extended states. The disorder potential is modeled by the on-site white-noise potential $V_{im}$ ($i$ denotes the column index, $m$ denotes the chain index) ranging from $-W/2$ to $W/2$. The effect magnetic field appears in the complex phase of the hopping term. The strength of the magnetic field is characterized by the flux per plaquette ($\phi$) in unit of magnetic flux quanta ($\phi_o = hc/e$). The Hamiltonian of this system can be written as:

$$\mathcal{H} = \sum_i \sum_{m=1}^{M} V_{im} |im\rangle\langle im| \qquad (2)$$
$$+ \sum_{\langle im;jn \rangle} \left[ t_{im;jn} |im\rangle\langle jn| + t_{im;jn}^\dagger |jn\rangle\langle im| \right],$$

where $<im;jn>$ indicates nearest neighbors on the lattice. The amplitude of the hopping term is chosen as the unit of the energy. For a specific energy $E$, a transfer matrix $T_i$ can be easily set up mapping the wavefunction amplitudes at column $i-1$ and $i$ to those at column $i+1$, i.e.

$$\begin{pmatrix} \psi_{i+1} \\ \psi_i \end{pmatrix} = T_i \begin{pmatrix} \psi_i \\ \psi_{i-1} \end{pmatrix} = \begin{pmatrix} H_i - E & -I \\ I & 0 \end{pmatrix} \begin{pmatrix} \psi_i \\ \psi_{i-1} \end{pmatrix}, \quad (3)$$

where $H_i$ is the Hamiltonian for the $i$th column, $I$ is a $M \times M$ unit matrix. Using a standard iteration algorithm [14], we can calculate the Lyapunov exponents for the transfer matrix $T_i$. The localization length $\lambda_M(E)$ for energy $E$ at finite width $M$ is then given by the inverse of the smallest Lyapunov exponent. In our numerical calculation, we choose the sample length to be over $10^4$ so that the self-averaging effect automatically takes care of the ensemble statistical fluctuations. We use the standard one-parameter finite-size scaling analysis [15] to obtain the thermodynamic localization length $\xi$. The scaling calculations are carried out with varying disorder $W$ at fixed energies. According to the one-parameter scaling theory, the renormalized finite-size localization length $\lambda_M/M$ can be expressed in terms of a universal function of $M/\xi$, i.e.,

$$\frac{\lambda_M(W)}{M} = f\left(\frac{M}{\xi(W)}\right), \quad (4)$$

where $f(x) \propto 1/x$ in the thermodynamic limit ($M \to \infty$) for localized states while approaching a constant ($\sim 1$) when $\xi$ diverges.

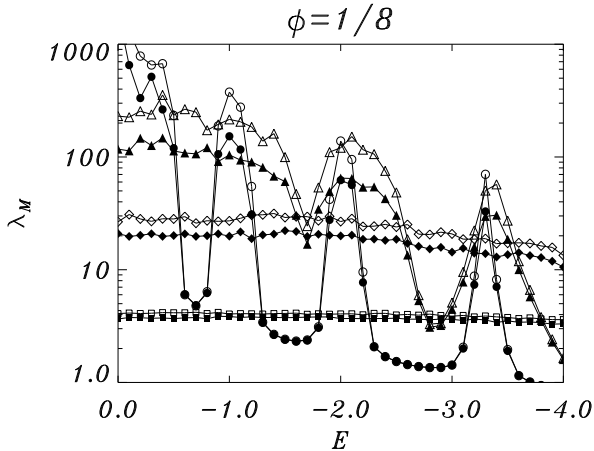

FIG. 1. Finite-size localization length ($\lambda_M$) as function of energy $E$ at magnetic field $\phi = 1/8$ for disorder $W = 1$ (circle), $W = 3$ (triangle), $W = 7$ (diamond) and $W = 12$ (square). The filled symbols are the data for system size $M = 32$ and the open symbols are for $M = 64$.

In Fig.1, we present the results of disorder $W$ dependence of finite-size localization length $\lambda_M$. Fig.1 shows the $\lambda_M$ at magnetic field $\phi = 1/8$ with system sizes $M = 32$ (filled symbols) and $M = 64$ (open symbols) for disorder strengths $W = 1$ (circle), $W = 3$ (triangle), $W = 7$ (diamond) and $W = 12$ (square). Due to the symmetry of the lattice model, only the lower energy branch results are shown here. In the case of weak disorder with $W = 1$, we can clearly see that the energy band does break up into small subbands (Landau bands). For a finite-size system, the states close to the Landau centers are extended since their localization lengths $\lambda_M$ depend on the system sizes as shown in Fig.1. We have calculated $\lambda_M$ for system sizes $M = 16, 24, 32, 48, 64$ and $84$. We found that $\lambda_M$ at the centers of Landau bands scale with the system size $M$. Therefore, the maxima in the finite-size localization length are the locations of the extended states even at thermodynamic limit. For those states between Landau bands, their localization lengths are independent of system size, and are localized states. We have done systematic studies with many diffrent $W$ values and found that as $W$ increases, the peak positions in $\lambda_M$ do not shift. Thus, we did not observe the floating-up of the extended state energy for increasing the disorder.

The transport property of a state is determined by the topological property-Chern number of the state [16,17]. For an eigenstate $|m\rangle$, the boundary phase averaged Hall conductance takes the form [16,17]

$$\langle \sigma_{xy}^m \rangle = \frac{1}{4\pi^2} \int d\theta_1 d\theta_2 \sigma_{xy}^m(\theta_1,\theta_2) = C(m) e^2/h, \quad (5)$$

where $C(m)$ is an integer called the Chern number of the state $|m\rangle$. States with nonzero Chern numbers are extended states which carry current and states with zero Chern numbers are the localized states. The transition for a state from being extended to localized is caused by the cancellation of Chern numbers and is equivalently a transition of Chern number from a non-zero integer to zero. For the case presented in Fig.1, the extended state close to the band center ($E = 0$) carries Chern number $C = -3$ while all the other extended states ($E \simeq -1.0, -2.0, -3.3$ in Fig.1) carry Chern numbers $C = +1$ [13,11,18]. As disorder strength $W$ increases, the $C = -3$ extended state moves toward the band bottom to cancel the $C = +1$ extended states one by one. Another possibility is for the $C = +1$ states to move toward band center to annihilate the $C = -3$ state which will correspond to the floating-up picture. However, the latter possibility is not consistent with our numerical findings. It is believed that extended state carrying non-zero Chern number can only move from $E_1$ to $E_2$ if there are no localized states between $E_1$ and $E_2$, or loosely speaking no mobility gap between $E_1$ and $E_2$. Fig.1 clearly confirmed this belief. As disorder increases from $W = 1$ to $W = 3$, all the states close to the band center $E = 0$ are extended, including those which are localized at $W = 1$. This can be seen by the separation in $\lambda_M$ for $M = 32$ and $M = 64$. However, those states close to the band bottom $E = -4$ are still showing Landau bands. At $W = 7$, the Landau band structure is totally gone and all states for a given system size $M$ have similar values for $\lambda_M$. Notice that in this case the $\lambda_M$ is of the same order as $M$, thus, all the states are some what extended in nature. Further increasing disorder to $W = 12$ will cause all the states to be localized and their $\lambda_M$ are much smaller than system size $M$. Above discussions about the cancellation of the Chern numbers is valid for a given finite system. In order to understand the transition in an infinite system, one has to study the thermodynamic localization length $\xi$ which is the focus of Fig.2 in the paper.

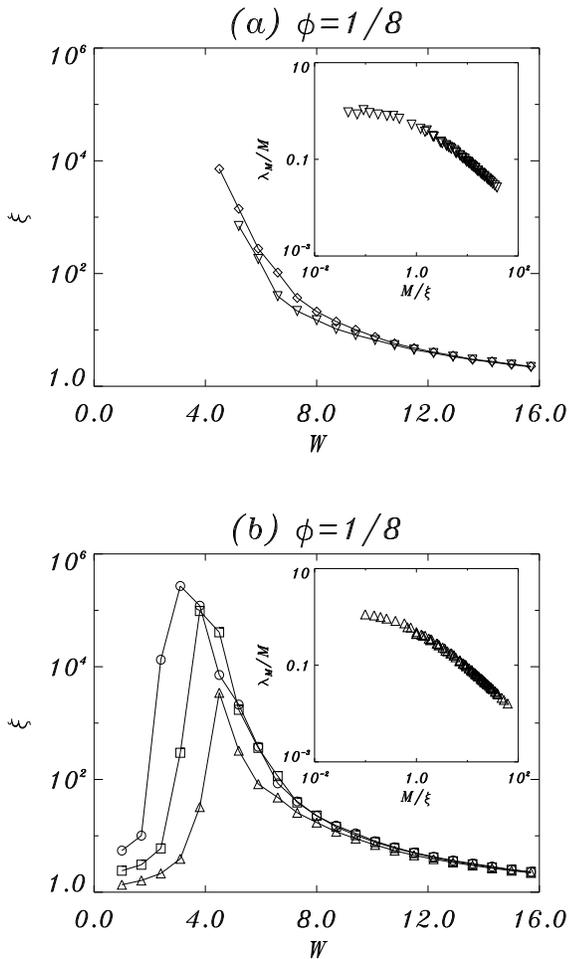

FIG. 2. Thermodynamic localization length $\xi$ for (a) $E = -2.0$ (diamonds) and $-3.3$ (down-triangles); and (b)$E = -0.7$ (circles), $-1.6$ (squares), and $-2.8$ (up-triangles). The scaling functions for $E = -3.3$ (a) and $E = -2.8$ (b) are shown in the insets.

In Fig.2 we present our results on thermodynamic localization length $\xi$ obtained from finite-size scaling calculations. To avoid the severe finite-size effect, we have used large system sizes in the scaling calculations with $M = 16, 24, 32, 48, 64$ and $84$. We carry out the scaling analysis with varying disorder $W$ at fixed energy $E$. In the insets, we show the scaling functions $\frac{\lambda_M(W)}{M} = f\left(\frac{M}{\xi(W)}\right)$ for both $E = -3.3$(a) and $E = -2.8$(b). The scaling plots are quite similar for the other energies we have calculated. One can see that all the data fall onto one smooth scaling curve from which we obtain the thermodynamic localization length $\xi$. In Fig.2, we show $\xi$ for five different energies $E = -0.7$ (circles), $-1.6$ (squares), $-2.0$ (diamonds), $-2.8$ (up-triangles) and $-3.3$ (down-triangles).

Presented in Fig. 2(a), $E = -3.3$ and $-2.0$ are the energies at the centers of the lowest and the first Landau bands when the disorder is weak such that the Landau bands are well defined as shown by the circles in Fig.1. These states are extended with infinite $\xi$ for $W < W_c$ and become localized when $W > W_c$. The important conclusion here is that the transitions to the insulator states are *identical* for *both* Landau bands. Therefore, we clearly see the direct transition from high LL quantum Hall state to insulator which is not allowed according to the global phase diagram [10]. We found similar critical disorder $W_c$ ($\simeq 4.5$) for both Landau bands. However, we should mention that it is difficult to estimate $W_c$ for the following reasons. At first, the energy value for the Landau band center can not be determined very accurately. Secondly, $\xi$ only approaches infinity from one side whereas for the case we are going to discuss below where $\xi$ diverges from both sides.

Presented in Fig. 2(b), $E = -0.7, -1.6$ and $-2.8$ are localized states at weak disorder limit as shown by the circles in Fig.1. These energy values are chosen because each of them is between two adjacent Landau bands and they are the most localized states in the valley. As $W$ increases, these states become more "extended" with increasing $\xi$. At critical disorder strengths $W_c$, these states become truly extended with divergent $\xi$. Beyond $W_c$, they become localized again. The critical strengths $W_c$ are different for different energies and it is smaller for the energy closer to the band center($E = 0$). Our best estimations are $W_c \simeq 3.5, 4.0$ and $4.5$ for $E = -0.7, -1.6$ and $2.8$. As we have mentioned above, the insulator transition is caused by the cancellation of the Chern number, and Chern number can only move from one energy to another if there is no mobility gap in between. The significance of the critical $W_c(E)$ is to allow the negative Chern number around band center $E = 0$ to move beyond $E$, thus, to annihilate the $+1$ Chern number state on the other side. Once the $+1$ Chern number is annihilated to zero, that quantum Hall state is transformed to an insulator. Therefore, because the difference in $W_c$ for different energies, we conclude that less disorder is required to induce higher LL to insulator transition.

Before conclusion, we would like to comment on the validity of the lattice model in describing real experiments. All the solid state systems should be described by lattice models. However, if the relevant length scale is much larger than the lattice spacing $a$, then a continuous model may be applicable. In a quantum Hall sample, the magnetic length $\ell$ is the relevant length which, in general, is much larger than the lattice spacing. On the other hand, there is an intrinsical difference between continuous and lattice models in this problem in the sense that there is no negative Chern number state at finite energy in a continuous model. Therefore, most likely, $E_c$ will float up in a continuous model as described in Eq.(1). The natural consequence is the global phase diagram [10] which predicts that there exists no direct phase transition between high LL quantum Hall state and insulator state. Clearly, this type of transition is allowed in the lattice model discussed here. In the previous work by us and Niu [11] with the same lattice model, we found that essential physics of the lattice model prevails even at very small field with large $\ell/a$. Recent experiment [6] might have already observed such transitions between high LL quantum Hall states and insulator states. We believe that the lattice model is necessary to understand this type of experiments for the reason that mentioned above.

After the work was completed, we became aware of a recent work by Yang and Bhatt [19]. They obtained similar numerical results as ours, however, their interpretation of the results are quite different. They conclude that although the positions of the extended states do not change much as were found in our work, the band position moves down as disorder $W$ increases. Therefore, relative to the band bottom, the positions for the extended states float up. There are two problems with this argument: (i) There have been calculations [11] with changing magnetic field at fixed disorder. In that case, band bottom does not change and, even there, the floating-up of $E_c$ is not seen. (ii) According to the original floating-up picture put forward by Laughlin and Khmelnitskii, the floating up only happens when disorder $W$ is of the order of Landau level spacing. However, using the reasonings by Yang and Bhatt, one would say that there is floating-up even for infinitesimal increasing in disorder $W$ which is quite different from the conventional understanding of the floating-up picture.

In summary, by calculating numerically the localization length in a lattice model, we studied the disorder induced transition between quantum Hall state and insulator state. We find the positions of the extended states do not change with varying disorder strength. As a consequence, there is direct transition from high Landau level quantum Hall state to insulator state, in contrast to the global phase diagram based on continuous models. This transition can be understood from the cancellation of the topological Chern numbers.

We thank S. Das Sarma, S. He, J.K. Jain, and especially Q. Niu and D.N. Sheng for many helpful discussions. D.Z. Liu is supported by NSF-DMR-91-20000 through the Science and Technology Center of Superconductivity.